\documentclass[10pt, conference, letterpaper]{IEEEtran}

\usepackage{algorithm}
\usepackage{algorithmicx}
\usepackage{algpseudocode}
\usepackage{amsfonts}
\usepackage{amsmath}
\usepackage{amssymb}
\usepackage[ansinew]{inputenc} 
\usepackage{xcolor}
\usepackage{mathtools}
\usepackage{graphicx}
\usepackage{caption}
\usepackage{subcaption}
\usepackage{import}
\usepackage{multirow}
\usepackage{cite}
\usepackage[export]{adjustbox}
\usepackage{breqn}
\usepackage{mathrsfs}
\usepackage{acronym}
\usepackage[keeplastbox]{flushend}
\usepackage{setspace}
\usepackage{stackengine}

\renewcommand{\thetable}{\arabic{table}}

\DeclareMathOperator*{\argmax}{arg\,max}

\def\delequal{\mathrel{\ensurestackMath{\stackon[1pt]{=}{\scriptscriptstyle\Delta}}}}

\graphicspath{{./figures/}}
\newcommand{\eq}[1]{Eq.~\eqref{#1}}
\newcommand{\fig}[1]{Fig.~\ref{#1}}

\newcommand{\secref}[1]{Section~\ref{#1}}

\renewcommand{\baselinestretch}{0.965}

\title{Beam Training and Data Transmission Optimization \\ in Millimeter-Wave Vehicular Networks}

\author{Maria Scalabrin$^\dag$, Nicol\`{o} Michelusi$^\ddag$, and Michele Rossi$^\dag$
\thanks{$^\dag$Dept.\ of Information Engineering, University of Padova, email: \{scalabri, rossi\}@dei.unipd.it}
\thanks{$^\ddag$School of Electrical and Computer Engineering, Purdue University, email: michelus@purdue.edu}\thanks{This research has been funded by NSF under grant CNS-1642982.}
} 

\IEEEoverridecommandlockouts

\begin{document}

\maketitle

\begin{abstract}
Future vehicular communication networks call for new solutions to support their capacity demands, by leveraging the potential of the \mbox{millimeter-wave} (\mbox{mm-wave}) spectrum.
Mobility, in particular, poses severe challenges in their design, and as such shall be accounted for. A key question in \mbox{mm-wave} 
vehicular networks is how to optimize the
\mbox{trade-off} between directive Data Transmission (DT) and directional Beam Training (BT), which enables it. In this paper, learning tools are investigated to optimize this \mbox{trade-off}. In the proposed scenario, a Base Station (BS) uses BT to establish a \mbox{mm-wave} directive link towards a Mobile User (MU) moving along a road. To control the BT/DT \mbox{trade-off}, a Partially Observable (PO) Markov Decision Process (MDP) is formulated, where the system state corresponds to the position of the MU within the road link. The goal is to maximize the number of bits delivered by the BS to the MU over the communication session, under a power constraint. The resulting optimal policies reveal that adaptive BT/DT procedures significantly outperform \mbox{common-sense} heuristic schemes, and that specific mobility features, such as user position estimates, can be effectively used to enhance the overall 
system performance and optimize the available system resources.
\end{abstract}
%
%


\section{Introduction}
\label{sec:Introduction}
The \mbox{state-of-the-art} protocols for vehicular communication address \mbox{vehicle-to-vehicle} (V2V) and \mbox{vehicle-to-infrastructure} (V2I) communication systems, generally termed V2X. Currently, these communication systems enable a maximum data rate of $100$~Mbps for high mobility (using 4G)~\cite{choi2016millimeter,giordani2017millimeter}, which are not deemed sufficient to support applications such as autonomous driving, augmented reality and infotainment, which will populate next-generation vehicular networks. Therefore, future vehicular communication networks call for new solutions to support their capacity demands, by leveraging the huge amount of bandwidth in the $30-300$~GHz band, the so called \mbox{millimeter-wave} (\mbox{mm-wave}) spectrum.
While communication at these frequencies is ideal to support high capacity demands, it relies on highly directional transmissions, which are extremely susceptible to the vehicle mobility. Therefore, a key question is: \emph{How do we leverage mobility information to optimize the \mbox{trade-off} between directive Data Transmission (DT) and directional Beam Training (BT), which enables it, to optimize the communication performance? How much do we gain by doing so?} 
To address these questions and optimize this \mbox{trade-off}, in this paper we envision the use of learning tools. We demonstrate significant gains compared to \mbox{common-sense} beam alignment schemes.

Compared to  conventional lower frequencies,
propagation at mm-waves poses several challenges, such as high propagation loss and sensitivity to blockage. To counteract these effects, mm-wave systems are expected to use large antenna arrays
to achieve a large beamforming gain via directional transmissions.
However, these techniques demand extensive beam training,
such as beam sweeping, estimation of angles of arrival and of departure, and \mbox{data-assisted} schemes~\cite{michelusi2018optimal}, 
as well as beam tracking~\cite{palacios2017tracking}. 
Despite their simplicity, the overhead incurred by these algorithms may ultimately offset the benefits of beamforming in highly mobile environments~\cite{choi2016millimeter,giordani2017millimeter}. 
While wider beams require less beam training, they result in a lower beamforming gain, hence smaller achievable capacity~\cite{7744807}. 
 While contextual information, such as GPS readings of vehicles~\cite{va2017inverse}, 
may alleviate this overhead, it does not eliminate the need for beam training due to noise and GPS acquisition inaccuracy. 
Thus, the design of schemes that alleviate this overhead is of great importance.
 
  In all of the aforementioned works, a priori information on the vehicle's mobility is not leveraged in the design of BT/DT protocols. 
  In contrast, \emph{we contend that leveraging such information via adaptive beam design techniques can greatly improve the performance of automotive networks}~\cite{va2015beam,va2016beam}.
 In this paper, we bridge this gap by designing adaptive strategies for BT/DT that leverage a priori mobility information via Partially Observable (PO) Markov Decision Processes (MDPs). Our numerical evaluations demonstrate that these optimized policies significantly outperform \mbox{common-sense} heuristic schemes, which are not tailored to the vehicle's observed mobility pattern.
Compared to~\cite{michelusi2018optimal}, which develops an
 analytical framework to  optimize
    the BT/DT \mbox{trade-off} and the BT parameters based on
   the "worst-case" mobility pattern, in this work, we assume a statistical mobility model.

In the proposed scenario, a Base Station (BS) attempts to establish a \mbox{mm-wave} directive link towards a Mobile User (MU) moving along a road. To this end, it alternates between BT and DT. 
The goal is to maximize the number of bits delivered by the BS to the MU over the communication session, under a power constraint.
To manage the BT/DT \mbox{trade-off}, we exploit a POMDP formulation, where the system state corresponds to the position of the MU within the road link. 
%
 Specifically, we implement a POMDP with temporally extended actions (i.e., actions with different durations) to model the different temporal scales of BT and DT, and a constraint on the available resources of the system. 
 POMDPs model an agent decision process in which the system dynamics are determined by the underlying MDP (in this case, the MU dynamics), but the agent cannot directly observe the system state. Instead, it maintains a probability distribution (called \emph{belief}) over the world states, based on observations and their distribution, and the underlying MDP. An exact solution to a POMDP yields the optimal action for each possible belief over the world states. 
POMDPs have been successfully implemented in a variety of \mbox{real-world} sequential decision processes, including robot navigation problems, machine maintenance, and planning under uncertainty 
 \cite{KAELBLING199899,kaelbling1996reinforcement}. 
  To address the complexity of POMDPs, we use ${\rm PERSEUS}$~\cite{DBLP:journals/corr/abs-1109-2145},
an  approximate solution technique which uses a \mbox{sub-set} of belief points as representative of the belief state. However, in contrast to the original formulation using random belief point selection, we tailor it by selecting a deterministic set of belief points representing 
uncertainty in MU position, and demonstrate significant performance gains.
A unified approach for constrained MDP is given by 
\cite{altman1999constrained,altman1995linear}. Notably, there has been relatively little development in the literature for incorporating constraints into the POMDP~\cite{isom2008piecewise,kim2011point,poupart2015approximate,undurti2010online}.
In order to address the resource constraints in our problem, we propose a Lagrangian method, and an online algorithm to optimize the Lagrangian variable based on the target cost constraint.

This paper is organized as follows. In~\secref{sec:System_Model}, we introduce the system model, followed by the optimization in~\secref{sec:Problem_Optimization}. We present numerical results in~\secref{sec:Numerical_Results}, followed by concluding remarks in~\secref{sec:Conclusions}.


\section{System Model}
\label{sec:System_Model}
We consider a scenario where a BS aims at establishing a \mbox{mm-wave} directive link with a MU moving along a road. To this end, it alternates between BT and DT: with BT, the BS refines its knowledge on the position of the MU within the road link, to perform more directive DT. Our goal is to maximize the number of bits that the BS delivers to the MU during a transmission episode, defined as the time interval between the two instants when the MU enters and exits the coverage range of the BS, under a power constraint.

\subsection{Problem formulation}
\label{sub:Problem_formulation}

\begin{figure}	
	\centering
	\includegraphics[width=0.65\columnwidth]{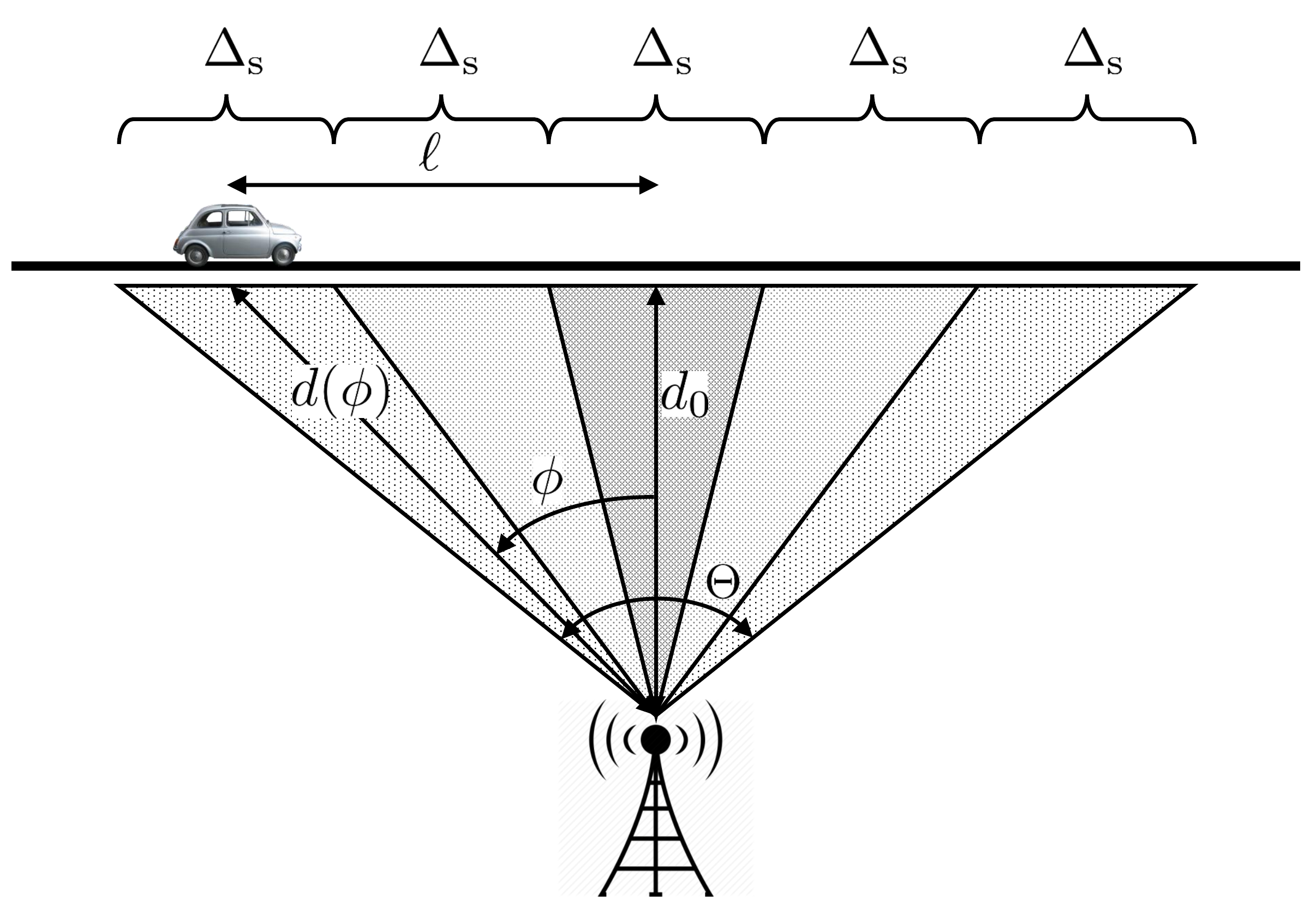}
	\caption{A \emph{dense} cell deployment.}\label{figure:Fig_scenario}
\end{figure}

We consider a \emph{dense} cell deployment, as shown in~\fig{figure:Fig_scenario}. The MU is associated with its closest BS, at a distance $d_0$ from the road link. The road link served by the reference BS is divided into $S$ road \mbox{sub-links} of equal length $\Delta_{\rm s} = 2d_0 \tan(\Theta/2)/S$, where $\Theta$ is the maximum coverage range of the BS. We let $\mathcal S\equiv\{1,\dots,S\}$ be the set of indices of the $S$ road \mbox{sub-links}. The BS associates a beam with each one of the $S$ road \mbox{sub-links}, with angular support, for the $s$-th beam,
\begin{equation}
\begin{split}
\!\!\Phi_s=\bigg[ &\tan^{-1}\frac{-d_0\tan(\Theta/2)+(s-1)\Delta_{\rm s}}{d_0}, \\
&\tan^{-1}\frac{-d_0\tan(\Theta/2)+s\Delta_{\rm s}}{d_0}\bigg]
\end{split}
\end{equation}
  and beamwidth $\theta_s=|\Phi_s|$, so that
  $\cup_{s\in\mathcal S} \Phi_s =[-\Theta/2,\Theta/2]$ and
   $\sum_{s\in\mathcal S} \theta_s = \Theta$. 

The time is discretized into micro \mbox{time-slots} of duration $\Delta_{\rm t}$, with $\Delta_{\rm t}$ being the time for a Primary Synchronization Signal (PSS), which allows a proper channel estimation at the receiver~\cite{7744807}.
At time $t$, the MU is located in one of the $S$ road \mbox{sub-links}, until it exits the coverage area of the BS, denoted by the absorbing state $\bar{s}=S+1$. We denote the \mbox{sub-link} occupied by the MU at time $t$ as $X_t\in\mathcal S$.
 We assume that the position of the MU within the road link evolves among the $S$ road \mbox{sub-links} following a random walk with probabilities $0<q<p<1$,
 with $1-p-q>0$,
 where $p=\mathbb P[X_{t+1}=s+1\mid X_t=s]$ and $q=\mathbb P[X_{t+1}=s-1\mid X_t=s]$. Under this model, the MU will exit the BS coverage area at some point.
 We can view such random walk as an abstraction of the following \emph{physical} mobility model, where the MU moves with
 average speed $\mathbb{E}[v]$ and speed variance $\text{Var}[v]$:
 assume that the MU moves at speed $v_t$ at time $t$, with
 $v_t \in \{0,v_{\rm max},-v_{\rm max}\}$. Also, let $\mathbb P[v_t=v_{\rm max}]=p$, $\mathbb P[v_t=-v_{\rm max}]=q$, and $\mathbb P[v_t=0]=1-p-q$. Note that the maximum speed supported by this model is  $v_{\rm max}\leq\Delta_{\rm s}/\Delta_{\rm t}$ (otherwise, the MU may move more than one \mbox{sub-link} within a single micro-slot). It follows that $\mathbb{E}[v] =v_{\rm max} (p-q)>0$ and $\text{Var}[v] =v_{\rm max}^2 (p+q) - (\mathbb{E}[v])^2>0$. Thus, 
 given average $\mathbb{E}[v]$ and  $\text{Var}[v]$, we obtain 
 $p$ and $q$ as
\begin{equation} \label{eq:pq}
\begin{split}
p & =\frac{\text{Var}[v]+(\mathbb{E}[v])^2}{2 v_{\rm max}^2} + \frac{\mathbb{E}[v]}{2 v_{\rm max}}, \\
q & =\frac{\text{Var}[v]+(\mathbb{E}[v])^2}{2 v_{\rm max}^2} - \frac{\mathbb{E}[v]}{2 v_{\rm max}} \, .
\end{split}
\end{equation}
To meet the conditions for the probabilities $0<q<p<1$, with $1-p-q>0$, the following inequalities must hold:
\begin{equation} \label{eq:pq_conditions}
\begin{split}
& p < 1 \to \text{Var}[v] < - (\mathbb{E}[v])^2 - v_{\rm max}\mathbb{E}[v] + 2 v_{\rm max}^2, \\
& q > 0 \to \text{Var}[v] > - (\mathbb{E}[v])^2 + v_{\rm max}\mathbb{E}[v], \\
&1-p-q > 0 \to \text{Var}[v]+(\mathbb{E}[v])^2<v_{\rm max}^2,
\end{split}
\end{equation} 
which defines a region of feasible pairs $(\mathbb{E}[v],\text{Var}[v])$. This model can be extended, e.g., to account for multiple speeds and memory in the velocity process, although we leave it for future work.

During BT or DT, at time $t$, the BS transmits 
using a beam that covers 
a \mbox{sub-set} of \mbox{sub-links}, $\hat{\mathcal S}_t\subseteq\mathcal S$, part of our design.
 Assuming a large antenna array, which allows for arbitrarily sharp beam patterns,
the beam is designed in order to support a target SNR ${\rm SNR}_t$ on the beam support,
$\mathcal B_t\equiv\cup_{s\in\hat{\mathcal S}_t}\Phi_s$.
To this end, we let $P_t(\phi)$ be the power per radian projected in the angular direction $\phi\in\mathcal B_t$, and $P_t(\phi)=0,\phi\notin\mathcal B_t$.
To attain the target SNR constraint, we must have that
\begin{align}
\frac{\Gamma P_t(\phi)}{
d(\phi)^2}={\rm SNR}_t,
\end{align}
where 
$\Gamma \delequal \lambda^2 \xi / (8\pi N_0 W_{\rm tot})$ is the SNR scaling factor, $\lambda=f_c/c$ is the wavelength, $N_0$ is the noise power spectral density, $\xi$ is the antenna efficiency,  $W_{\rm tot}$ is the bandwidth, and
 $d(\phi)=d_0\sqrt{1+\tan(\phi)^2}$ is the distance of the point in the road link at angular direction $\phi$, so that $d(\phi)^{-2}$ models distance dependent path loss.
It follows that the total transmit power is given by
\begin{align} 
\!\!P_t=\int_{\mathcal B_t}P_t(\phi){\rm d}\phi
={\rm SNR}_t
\sum_{s\in\hat{\mathcal S}_t}
\int_{\Phi_s}\frac{d_0^2}{\Gamma}[1+\tan(\phi)^2]{\rm d}\phi.
\end{align}
Using the change of variables
$\phi \to \ell = d_0\tan(\phi)$, we then obtain
\begin{align} \label{eq:P_t_2}
P_t
={\rm SNR}_t\frac{1}{\Gamma}\sum_{s\in\hat{\mathcal S}_t}
\int_{-d_0\tan(\Theta/2)+(s-1)\Delta_{\rm s}}
^{-d_0\tan(\Theta/2)+s\Delta_{\rm s}}\!\!\!\!\!\!\!\!\!d_0{\rm d}\ell
=
{\rm SNR}_t\frac{\Delta_{\rm s}d_0}{\Gamma}|\hat{\mathcal S}_t|.
\end{align}
In other words, the total transmit power is independent of the \mbox{sub-link} indices,
but depends solely on the number of \mbox{sub-links} $|\hat{\mathcal S}_t|$ and on the target SNR. This result is in line with the intuition that larger distances are achievable via smaller beamwidths, and vice versa~\cite{7744807}.

During DT, assuming isotropic reception at the MU,
 such target SNR implies an achievable rate given by
\begin{equation} \label{eq:R_t}
R_t = W_{\rm tot} \log_2 \bigg (1+
\frac{\Gamma}{\Delta_{\rm s}d_0}\frac{P_t}{|\hat{\mathcal S}_t|}
\bigg ) \, .
\end{equation}


During BT, the SNR is set so as to achieve target \mbox{mis-detection} and \mbox{false-alarm} probabilities.
To design this parameter, the generic signal detection problem corresponds to receiving a signal $y[l]$, $l=1,\dots,L$, over a noisy channel. The two hypotheses are
\begin{equation} \label{eq:SD_1}
\begin{split}
& \mathcal{H}_0 : y[l]=w[l] \qquad \text{ (no signal at the RX)}\\
& \mathcal{H}_1 : y[l]=x[l]+w[l] \quad \text{(signal at the RX)}\\
\end{split}
\end{equation}
where $w[l]$, $l=1,\dots,L$, are independent random variables, $w[l]\sim\mathcal{N}(0,\sigma_w^2)$, with $\sigma_w^2=N_0$. Our task is to decide in favor of $\mathcal{H}_0$ or $\mathcal{H}_1$ on the basis of the measurements $y[l]$, $l=1,\dots,L$, i.e., 
\begin{equation*}
\mathbb P(y[1],\dots,y[L] | \mathcal{H}_1) \mathbb P(\mathcal{H}_1) \gtrless \mathbb P(y[1],\dots,y[L] | \mathcal{H}_0) \mathbb P(\mathcal{H}_0),
\end{equation*}
or equivalently,
\begin{equation} \label{eq:SD_3}
\sum_{l=1}^L y[l]x[l] \gtrless \sigma_w^2 \ln \bigg ( \frac{\mathbb P(\mathcal{H}_0)}{\mathbb P(\mathcal{H}_1)} \bigg ) + \frac{1}{2}E_x \, ,
\end{equation}
where $E_x=\sum_{l=1}^L x[l]^2$ is the energy of the pilot signal $x[l]$. If the \mbox{Neyman-Pearson} formulation is used, then the right hand side of~\eq{eq:SD_3} is 
replaced by a decision threshold $\bar{\rho}$, function of the target error probability. According to the \mbox{Neyman-Pearson} Lemma~\cite{kay2013fundamentals}, for a given target error probability, we can derive a decision rule as follows. The \mbox{false-alarm} probability, $\mathbb P_{\text{FA}}$ (accept $\mathcal{H}_1$ when $\mathcal{H}_0$ is true),
 is given as $\mathbb P_{\text{FA}}=\int_{\bar{\rho}}^\infty \mathbb P(y \mid \mathcal{H}_0) dy=Q(\frac{\bar{\rho}}{\sigma_w \sqrt{E_x}})$, where $Q(\cdot)$ is the \mbox{Q-function}.
The \mbox{mis-detection} probability, $\mathbb P_{\text{MD}}$ (accept $\mathcal{H}_0$ when $\mathcal{H}_1$ is true),
is given as $\mathbb P_{\text{MD}}=1-\mathbb P_{\text{D}}$, where the probability of correct detection is given by $\mathbb P_{\text{D}}=\int_{\bar{\rho}}^\infty \mathbb P(y \mid \mathcal{H}_1) dy=Q(\frac{\bar{\rho}-E_x}{\sigma_w \sqrt{E_x}})=Q(Q^{-1}(\mathbb P_{\text{FA}}) - \frac{\sqrt{E_x}}{\sigma_w})$, which shows that $P_{\text{D}}$ is a function of $\mathbb P_{\text{FA}}$. Applying the inverse $Q^{-1}(\cdot)$ to both sides of the last equation, leads to a measure of the SNR required to attain the target error performance: 
\begin{equation} \label{eq:target_SNR}
{\rm SNR}_t=\frac{E_x}{\sigma_w^2}=(Q^{-1}(\mathbb P_{\text{FA}})-Q^{-1}(\mathbb P_{\text{D}}))^2,
\end{equation} 
which is plugged into \eq{eq:P_t_2} to find the transmit power as a function of the
number of \mbox{sub-links} covered, $|\hat{\mathcal S}_t|$.

\subsection{Partially Observable Markov Decision Process}
\label{sec:Partially Observable Markov Decision Process}

Next, we define a constrained Partially Observable (PO) Markov Decision Process (MDP). 

\noindent \textbf{States:} $\mathcal{\bar{S}}$ is a finite set of states describing the position of the MU within the road link, 
 along with the absorbing state $\bar s$ when the MU exits the coverage area of the BS.
 Therefore, $\mathcal{\bar{S}}\equiv\mathcal S\cup\{\bar s\}$,
 where
 $\mathcal S\equiv\{1,\dots,S\}$ is the set of road \mbox{sub-links}.

\noindent \textbf{Actions:} $\mathcal{A}$ is a finite set of actions that the BS can perform. Specifically, the BS can perform actions for Beam Training (BT) and actions for Data Transmission (DT),
which involve selection of the transmission beam, power, and duration.
In general, $a \in \mathcal{A}$ is in the form $a = (\hat{\mathcal S}, P_c, T_c)$, where: $c = \{ {\text{BT}}, {\text{DT}} \}$ refers to the action class; $\hat{\mathcal S}\subseteq\mathcal S$ is a \mbox{sub-set} of \mbox{sub-links}, defining the support of the transmission beam;
  $P_c$ is the transmission power per beam, such that $P_t=P_c |\hat{\mathcal S}|$ in \eq{eq:P_t_2} is the total transmit power,
   $T_c$ is the transmission duration of action $a \in \mathcal{A}$ (number of micro \mbox{time-slots} of duration $\Delta_{\rm t}$).
If $c={\text{BT}}$, then $a = (\hat{\mathcal S}, P_{\text{BT}}, T_{\text{BT}})$, where $P_{\text{BT}}$ and $T_{\text{BT}}$ are fixed parameters of the model. Specifically, we assume that BT actions perform simultaneous beamforming over $\hat{\mathcal S}$ in one interval of $\Delta_{\rm t}$ seconds (i.e., $T_{\text{BT}}=1$). Also, $P_{\text{BT}} = P_{\rm min}$, where $P_{\min}$ is the power per beam required to attain the target SNR constraint, i.e., $P_{\min}={\rm SNR}\frac{\Delta_{\rm s}d_0}{\Gamma}$, and ${\rm SNR}$ is a function of false-alarm and mis-detection probabilities $\mathbb P_{\text{FA}}$ and $\mathbb P_{\text{MD}}$, which are also fixed parameters of the model, via \eq{eq:target_SNR}.
If $c={\text{DT}}$, then $a = (\hat{\mathcal S}, P_{\text{DT}}, T_{\text{DT}})$, where $P_{\text{DT}}$ and $T_{\text{DT}}$ are part of the optimization.  Specifically, we assume that DT actions perform simultaneous data communication over $\hat{\mathcal S}$  for $T_{\text{DT}}-1$ micro time-slots, where the last interval of $\Delta_{\rm t}$ seconds is dedicated to the ACK/NACK feedback transmission from the MU to the BS. During DT, the transmission rate follows from \eq{eq:R_t}.
 Note that the action space 
 grows as $|\hat{\mathcal S}|= \sum_{s=1}^{S} S!/s!/(S-s)! = 2^{S}-1$. To reduce its cardinality, we restrict $\mathcal{A}$ such that $\hat{\mathcal S}$ is a \mbox{sub-set} of {\it consecutive} indices in $\mathcal{S}$, i.e., the beam directions specified by $\hat{\mathcal S}$ define a compact range of transmission for the BS. Thus, $|\hat{\mathcal S}|=S(S+1)/2$. \\
\noindent \textbf{Observations:} $\mathcal{O}$ is a finite set of observations,
 defined as $\mathcal O\equiv\{ {\text{ACK}}, {\text{NACK}}, \bar{s} \}$. Specifically, $o=\bar{s}$ means that
the MU exited the coverage area of the BS; for simplicity, in this work we assume that such event is observable, i.e., the BS knows when the MU exited its coverage area.

\noindent \textbf{Transition probabilities:} 
$\mathbb P(s'|s,a)$ is the transition probability from $s\in\mathcal{\bar{S}}$ to $s^\prime\in\mathcal{\bar{S}}$ 
given $a \in \mathcal{A}$. Note that these probabilities are 
a function of the duration $T_c$ of action $a$. 
If the transmission duration of $a \in \mathcal{A}$ is $T_c=1$, then we store the \mbox{$1$-step} transition probabilities into matrix $\mathbf M=[\mathbf M_{ss'}]$, with elements $\mathbf M_{ss'}=\mathbb P(s'|s,a)$ given by the 1-step mobility model, as:
\begin{equation} \label{eq:T}
\begin{aligned}
& \mathbb P(s'|s,a) = p &\text{$s'=s+1$, $s=1,\dots,S$} \\
& \mathbb P(s'|s,a) = q &\text{$s'=s-1$, $s=2,\dots,S$} \\
& \mathbb P(s'|s,a) = 1-p-q &\text{$s'=s$, $s=2,\dots,S$} \\ 
& \mathbb P(s'|s,a) = 1-p &\text{$s'=s$, $s=1$} \\ 
& \mathbb P(s'|s,a) = 1 &\text{$s'=s$, $s=\bar{s}$.} \\ 
\end{aligned}
\end{equation}
If the transmission duration of $a \in \mathcal{A}$ is $T_c=N$, then we compute the \mbox{$N$-step} transition probabilities into matrix $\mathbf M^{N}$, i.e., we take the $N$-th power of matrix $\mathbf M$ 
so as to account for the \mbox{$N$-step} evolution of the system state under $a \in \mathcal{A}$ with transmission duration $T_c$. 

\noindent \textbf{Observation model:} $\mathbb P(o|s,a,s')$ is the probability of observing $o \in \mathcal{O}$ given $s \in \mathcal{\bar{S}}$ and $a \in \mathcal{A}$ with transmission duration $T_c$, ending in $s' \in \mathcal{\bar{S}}$. We assume that the BS can successfully perform $a = (\hat{\mathcal S}, P_c, T_c)$ if the MU remains within $\hat{\mathcal S}$ for $T_c$ subsequent micro \mbox{time-slots}, i.e., the MU does not exit from the beam support, so that all signal is received. In this case, the MU feeds back an ACK to the BS, $o=\text{ACK}$. Therefore, we define $X_0^{T_c}=\{X_0, \dots, X_{T_c}\}$, as the system state path from time $0$ to time $T_c$, and the event $E_{s,s'}^{N} = \{ X_0^{T_c} \in \hat{\mathcal S}^{T_c+1}\mid X_0=s, X_{T_c}=s', T_c=N \}$, meaning that the system state path $X_0^{T_c}$ remains within $\hat{\mathcal S}$ for $T_c$ subsequent micro \mbox{time-slots}, given that $X_0=s$, $X_{T_c}=s'$, $T_c=N$. 
In order to compute it, we also define matrix $\tilde{\mathbf M}$ 
as the transition probability matrix restricted to the beam support $\hat{\mathcal S}$, i.e.,
$\tilde{\mathbf M}=[\tilde{\mathbf M}_{ss'}]$, with elements $\tilde{\mathbf M}_{ss'}=\mathbf M_{ss'}$ if $\{s,s'\} \in \hat{\mathcal S}$, otherwise $\tilde{\mathbf M}_{ss'}=0$. We derive $\mathbb P(E_{s,s'}^{N})$ as: 

\begin{align} \label{eq:P_E}
\nonumber
&\mathbb P(E_{s,s'}^{N}) = \mathbb P( X_0^{T_c} \in \hat{\mathcal S}^{T_c+1}\mid X_0=s, X_{T_c}=s', T_c=N) \\
\nonumber
&\!\!\!=\!\!\frac{1}{\mathbf M_{ss'}^{N}} \mathbb P(X_0^{T_c} \in \hat{\mathcal S}^{T_c+1}, X_{T_c}=s'\mid X_0 = s, T_c=N)\\
&\!\!\!=\!\!\frac{1}{\mathbf M_{ss'}^{N}}\!\!\sum_{s_0^{N} \in \hat{\mathcal S}^{N+1}} \prod_{z=0}^{N-1} \mathbb P(X_{z+1}{=}s_{z+1}\mid X_z{=}s_z) {=}\frac{\tilde{\mathbf M}_{ss'}^N}{\mathbf M_{ss'}^{N}}.
\!\!
\end{align}


Given $a \in \mathcal{A}$ with transmission duration $T_c$,
$\mathbb P(o|s,a,s')$ is defined as follows. 
If $c={\text{BT}}$,
 we account for \mbox{false-alarm} and \mbox{mis-detection} errors in the beam detection process.
 In particular, if $\{s,s'\} \in\hat{\mathcal S}$ (i.e., the MU is within the beam support during the
 duration of BT) then $\mathbb P({\text{ACK}}|s,a,s')=\mathbb P_{\text{D}}$ (correct detection) and $\mathbb P({\text{NACK}}|s,a,s')=\mathbb P_{\text{MD}}$ (\mbox{mis-detection}); 
 on the other hand, if $\{s,s'\}\notin\hat{\mathcal S}$ (i.e., the MU is outside of the beam support during the duration of BT), then 
 $\mathbb P({\text{ACK}}|s,a,s')=\mathbb P_{\text{FA}}$ (\mbox{false-alarm}) and $\mathbb P({\text{NACK}}|s,a,s')=1-\mathbb P_{\text{FA}}$.
If $c={\text{DT}}$, then the transmission is successful if the event $E_{s,s'}^{N}$ occurs,
 so that $\mathbb P({\text{ACK}}|s,a,s')=\mathbb P(E_{s,s'}^{N})$ for $\{s,s'\}\in\mathcal{S}$, and $\mathbb P({\text{NACK}}|s,a,s')=1-\mathbb P({\text{ACK}}|s,a,s')$.
 Finally, 
  $\mathbb P(\bar{s}|s,a,s')=1$ whenever either $s=\bar{s}$ or $s'=\bar{s}$, i.e., the BS knows when the MU exited its coverage area. 

\noindent \textbf{Rewards:} $r(s,a)$ is the expected reward given $s \in \mathcal{\bar{S}}$ and $a \in \mathcal{A}$, defined as the transmission rate (number of bits transmitted from the BS to the MU) during DT 
if the MU remains within $\hat{\mathcal S}$ for $T_c$ subsequent micro \mbox{time-slots}. Formally, $r(s, a)=\sum_{s' \in \mathcal{\bar S}} \mathbb P(s' | s,a)r(s,a,s')=\sum_{s' \in \mathcal{\bar S}}\mathbf M_{ss'}^{N}\mathbb{E}[ R (T_{\text{DT}}-1) \mathcal{X}(E_{s,s'}^{N})]$, where $\mathcal{X}(E_{s,s'}^{N})=1$ iff the event $E_{s,s'}^{N}$ is true (thus $r(s,a)=0$ if the MU exits from the beam support). Note that $\mathbb{E}[\mathcal{X}(E_{s,s'}^{N})]= \mathbb P(E_{s,s'}^{N})$, which is computed in \eq{eq:P_E}. The transmission rate $R$ follows from \eq{eq:R_t} when $P_t = P_{\text{DT}}|\hat{\mathcal S}|$. Finally, $r(s, a) = R (T_{\text{DT}}-1) \sum_{s' \in \mathcal{\bar S} } \tilde{\mathbf M}_{ss'}^N$, where $T_{DT}-1$ refers to the fact that we reserve one micro \mbox{time-slot} over the total DT duration for the feedback transmission. If $c={\text{BT}}$, then $r(s, a) = 0$, as no bits of data are transmitted.

\noindent \textbf{Costs:} $c(s,a)$ is the expected energy cost given $s \in \mathcal{\bar{S}}$ and $a \in \mathcal{A}$. The total expected cost during a transmission episode is subject to the constraint $C$.
If $c=\text{DT}$, then $c(s,a)=P_{\text{DT}}|\hat{\mathcal S}|(T_{\text{DT}}-1)$, $\forall s \in \mathcal{S}$ (we reserve one micro \mbox{time-slot} for the feedback transmission). If $c={\text{BT}}$, then $c(s, a) = P_{\text{BT}}|\hat{\mathcal S}|$. 

%
%


\vspace{-2mm}
\section{Optimization Problem}
\label{sec:Problem_Optimization}

Since the agent cannot directly observe the system state, we introduce the notion of \emph{belief}. A belief $b \in \mathcal{B}$ is a probability distribution over $\mathcal{\bar{S}}$. The state estimator must compute a new belief, $b' \in \mathcal{B}$, given an old belief $b \in \mathcal{B}$, an action $a \in \mathcal{A}$, and an observation $o \in \mathcal{O}$, i.e., $b'=\tau(o,a,b)$. It can be obtained via Bayes' rule as:
\begin{equation}\label{eq:belief_update}
\begin{split}
b'(s') & = \mathbb P(s'\mid o, a, b) \\
& = \frac{ \sum_{s \in \mathcal{\bar{S}}} \mathbb P(o|s,a,s') \mathbb P(s'|s,a) b(s)}{\mathbb P(o|a,b)}.
\end{split}
\end{equation}
where $\mathbb P(o\mid a, b)$ is a normalizing factor, $\sum_{s' \in \mathcal{\bar{S}}} b'(s') = 1$.

Our goal is to determine a policy $\pi$ (i.e., a map from beliefs to actions) that maximizes the total expected reward the agent can gather,
under a constraint on the total expected cost during a transmission episode,
 following $\pi$ and starting from $b_0=b$:
\begin{equation} \label{eq:OPT}
\begin{split}
& \max_\pi \, \mathbb{E}_{\pi} \bigg [ \sum_{t=0}^\infty r_t(b_t, \pi(b_t)) \mid b_0 = b \bigg ], \\
& \quad \text{s.t. $\mathbb{E}_{\pi} \bigg [ \sum_{t=0}^\infty c_t(b_t, \pi(b_t)) \mid b_0 = b \bigg ] \le C$} \\
\end{split}
\end{equation}
where we have defined expected rate and cost metrics under belief $b_t$ as
\begin{equation}
\begin{split}
& r_t(b_t, \pi(b_t))=\sum_{s\in\mathcal{\bar S}}r_t(s, \pi(b_t))b_t(s) \\
& c_t(b_t, \pi(b_t))=\sum_{s\in\mathcal{\bar S}}c_t(s, \pi(b_t))b_t(s) \, .
\end{split}
\end{equation}


At this point, we opt for a Lagrangian relaxation approach such that $\mathcal{L}(s,a)=r(s,a)-\lambda c(s,a)$ is the metric to be maximized, for some Lagrangian multiplier $\lambda \ge 0$, and the total expected cost during a transmission episode is subject to the constraint $C$. 
Hereinafter, according to the notation, $\mathcal{L}_{t}(b_t, \pi(b_t))=\sum_{s\in\mathcal{\bar S}}\mathcal{L}_{t}(s, \pi(b_t))b_t(s)$.
At the end of \secref{sec:PERSEUS}, we will consider an online algorithm to optimize parameter $\lambda$ so as to solve the original problem in \eq{eq:OPT}.

\subsection{Value Iteration for POMDPs}
\label{sec:Value Iteration for POMDPs}

In POMDPs, a policy $\pi$ is a function over a continuous set $\mathcal{B}$ of probability distributions over $\mathcal{\bar{S}}$. A policy $\pi$ is characterized by a value function \mbox{$V^\pi:\mathcal{B} \to \mathbb{R}$}, which is defined as:
\begin{equation} \label{eq:V_b}
V^\pi(b) = \mathbb{E}_{\pi} \bigg [ \sum_{t=0}^\infty \mathcal{L}_{t}(b_t, \pi(b_t)) \mid b_0 = b \bigg ] \, .
\end{equation}
A policy $\pi$ that maximizes $V^\pi$ is called an optimal policy $\pi^*$. 
The value of an optimal policy $\pi^*$ is the optimal value function $V^*$, that satisfies the Bellman optimality equation $V^* = H V^*$ (with Bellman backup operator $H$):
\begin{equation} \label{eq:V_b_update}
 V^*(b){=}\max_{a \in \mathcal{A}} \bigg [\sum_{s \in \mathcal{\bar{S}}} b(s) \mathcal{L}(s, a) {+}\sum_{o \in \mathcal{O}} \mathbb P(o|a, b) V^*(b') \bigg ],
\end{equation}
where $b'=\tau(o,a,b)$ (see \eq{eq:belief_update}). When \eq{eq:V_b_update} holds for every belief $b \in \mathcal{B}$ we are ensured the solution is optimal. $V^*$ can be arbitrarily well approximated by iterating over a number of stages, at each stage considering a step further into the future. 
 Also, for problems with an infinite planning horizon, $V^*$ can be approximated, to any degree of accuracy, by a PieceWise Linear and Convex (PWLC) value function~\cite{DBLP:journals/corr/abs-1109-2145}. Thus, $V_{n+1}=HV_n$ and we parameterize a value function $V_n$ at stage $n$ by a finite set of vectors (hyperplanes) $\{\alpha_n^i \}_{i=1}^{|V_n|}$, such that $V_n(b)=\max_{\{\alpha_n^i \}_{i=1}^{|V_n|}} b \cdot \alpha_n^i$, where $(\cdot)$ denotes inner product. Each vector in $\{\alpha_n^i \}_{i=1}^{|V_n|}$, is associated with an action $a(\alpha_n^i) \in \mathcal{A}$, which is the optimal one to take at stage $n$, and defines a region in the belief space for which this vector is the maximizing element of $V_n$ (thus $\pi(b)=a(\alpha_n^i)$). The key idea is that for a given value function $V_n$ at stage $n$ and a belief $b \in \mathcal{B}$, we can compute the vector $\alpha_{n+1}^b$ in $\{\alpha_{n+1}^i \}_{i=1}^{|HV_n|}$ such that:
\begin{equation} \label{eq:alpha}
\alpha_{n+1}^b = \argmax_{\{\alpha_{n+1}^i \}_i^{|HV_n|}} b \cdot \alpha_{n+1}^i \, ,
\end{equation}
where $\{\alpha_{n+1}^i \}_{i=1}^{|HV_n|}$ is the (unknown) set of vectors for $HV_n$. We will denote this operation $\alpha_{n+1}^b = {\rm backup}(b)$. It computes the optimal vector for a given belief $b \in \mathcal{B}$ by back-projecting all vectors in the current horizon value function one step from the future and returning the vector that maximizes the value of $b \in \mathcal{B}$. Defining vectors $\mathcal{L}_a$ such that $\mathcal{L}_a(s) = \mathcal{L}(s, a)$ and $g_{a,o}^i$ such that $g_{a,o}^i(s) = \sum_{s' \in \mathcal{\bar S}} \mathbb P(o|s,a,s') \mathbb P(s'|s,a) \alpha_n^i(s')$ ($g_{a,o}^i$ is a projected vector given action $a$, observation $o$, and current horizon vector $\alpha_n^i$), we have~\cite{DBLP:journals/corr/abs-1109-2145}:
\begin{equation} \label{eq:backup}
\begin{split}
{\rm backup}(b) & = \argmax_{ \{ g_a^b \}_{a \in \mathcal{A}} } b \cdot g_a^b \\
& = \argmax_{ \{ g_a^b \}_{a \in \mathcal{A}} } b \cdot [ \mathcal{L}_a + \sum_{o \in \mathcal{O}} \argmax_{ \{ g_{a,o}^i \}_{i} } b \cdot g_{a,o}^i ] \, .
\end{split}
\end{equation}
In general, computing optimal planning solutions for POMDPs is an intractable problem for any reasonably sized task. This calls for approximate solution techniques, e.g., ${\rm PERSEUS}$~\cite{DBLP:journals/corr/abs-1109-2145}, which we introduce next.

\subsection{Randomized Point-based Value Iteration for POMDPs}
\label{sec:PERSEUS}

${\rm PERSEUS}$ is an approximate \mbox{Point-Based} Value Iteration (PBVI) algorithm for POMDPs. It implements a randomized approximate backup operator $\tilde{H}$ that increases (or at least does not decrease) the value of all beliefs $b \in \tilde{\mathcal{B}} \subset \mathcal{B}$. The key idea is that for a given value function $V_n$ at stage $n$, we can build a value function $V_{n+1} = \tilde{H} V_n$ that improves the value of all beliefs $b \in \tilde{\mathcal{B}} \subset \mathcal{B}$ by only updating the value of a (randomly selected) subset of beliefs $\tilde{\mathcal{B}} \subset \mathcal{B}$, i.e., we can build a value function $V_{n+1} = \tilde{H} V_n$ that upper bounds $V_n$ over $\tilde{\mathcal{B}} \subset \mathcal{B}$ (but not necessarily over $\mathcal{B}$): $V_n(b) \le V_{n+1}(b)$, $\forall b \in \tilde{\mathcal{B}} \subset \mathcal{B}$.
Starting with $V_0$, ${\rm PERSEUS}$ performs a number of backup stages until some convergence criterion is met. Each backup stage is defined as in Algorithm~\ref{alg:alg_1} (where $\tilde{\mathcal{B}}_{\rm temp}$ is an auxiliary set containing the non-improved beliefs).

\begin{algorithm}[h]
\caption{function {\rm PERSEUS}
\label{alg:alg_1}}
\begin{algorithmic}[1]
\Function{Perseus}{$\tilde{\mathcal{B}}$, $V_n$}
	\State Set $V_{n+1} = 0$. Initialize $\tilde{\mathcal{B}}_{\rm temp}$ to $\tilde{\mathcal{B}}$.
	\While{$\tilde{\mathcal{B}}_{\rm temp} \neq \emptyset$}
		\State Sample $b$ from $\tilde{\mathcal{B}}_{\rm temp}$ 
		\State Compute $\alpha = {\rm backup}(b)$ (see~\eq{eq:backup})
		\If{$b \cdot \alpha \ge V_n(b)$} 
			\State Add $\alpha$ to $V_{n+1}$
		\Else
			\State Add $\alpha' =  \argmax_{\{\alpha_n^i \}_i^{|V_n|}} b \cdot \alpha_n^i$ to $V_{n+1}$
		\EndIf
	\State Compute set $\tilde{\mathcal{B}}_{\rm temp} = \{b \in \tilde{\mathcal{B}} : V_{n+1}(b) < V_n(b) \}$
	\EndWhile
  	\State \Return $V_{n+1}$
\EndFunction
\end{algorithmic}
\end{algorithm}


\begin{algorithm}[tbh]
\caption{function BELIEFS
\label{alg:fun_BELIEFS}}
\begin{algorithmic}[1] 
\Function{Beliefs}{}
	\State $\tilde{\mathcal{B}}=\emptyset$
	\For{$w=1:W$}
		\For{$i=1:S+1-w$}
			\State Build $b$ such that $b_{i:i+w-1}=1/w$ and $\tilde{\mathcal{B}} \leftarrow b$
		\EndFor
	\EndFor
	\State \Return $\tilde{\mathcal{B}}$
\EndFunction
\end{algorithmic}
\end{algorithm}

Key to the performance of ${\rm PERSEUS}$ is the design of $\tilde{\mathcal{B}}$. 
Several standard schemes to select beliefs have been proposed for PBVI, mainly based on grids of points in the belief space. A different option to select beliefs is to simulate the model, i.e., sampling random actions and observations, and generating trajectories through the belief space, as suggested in~\cite{DBLP:journals/corr/abs-1109-2145}. Although this approach may seem reasonable, one may argue that the probability distributions collected in $\tilde{\mathcal{B}}$ are not very representative of the system dynamics history, where actions and observations must also depend on beliefs. Hereinafter, we leverage the structure of the POMDP presented in \secref{sec:System_Model} and provide an algorithm (Algorithm~\ref{alg:fun_BELIEFS}) to collect beliefs in $\tilde{\mathcal{B}}$ in a smarter fashion. Our approach is simple but effective, and does not require any prior knowledge of the system dynamics: according to Algorithm~\ref{alg:fun_BELIEFS}, $\tilde{\mathcal{B}}$ is made of uniform probability distributions over $\mathcal{\bar S}$, which are uniformly distributed over at most $W$ \emph{consecutive} road \mbox{sub-links}. Then, this design of $\tilde{\mathcal{B}}$ reflects the compact range of transmission for the BS, where the BS degree of uncertainty on the MU state scales with $W$.


The basic routine for PBVI is given in Algorithm~\ref{alg:PBVI}, where $V_{n+1}=V_{n+1}^r-\lambda V_{n+1}^c$ approximates the optimal value function for a given value of $\lambda$. Note that we are interested in the optimal policy $\pi^*$ when $b_0$ is such that $b_0(s)=\delta(s=1)$, i.e., the agent knows when the MU enters the maximum coverage range of the BS. 

\begin{algorithm}[tbh]
\caption{Point-Based Value Iteration (PBVI)
\label{alg:PBVI}}
\begin{algorithmic}[1] 
\State $\tilde{\mathcal{B}}=\text{BELIEFS}$
\State Set $n=0$, $V_0=0$, $V_0^c=0$, $\lambda_0={\lambda}$, $V_{\rm opt}=-\infty$.
\State Define $\mathcal{L}(s,a)=r(s,a)-\lambda_0 c(s,a)$
	\For{$n=0,\dots$}
		\State $V_{n+1} = \text{PERSEUS}(\tilde{\mathcal{B}},V_n)$
		\If{$|\sum_{b \in \tilde{\mathcal{B}}} V_{n+1}(b) / \sum_{b \in \tilde{\mathcal{B}}} V_n(b) -1|<\epsilon_V$}
			\State Break
		\EndIf
	\EndFor
	\State $V^*(b_0)=V_{n+1}(b_0)$
	\State $V_c^*(b_0)=V_{n+1}^c(b_0)$
	\If{$V_c^*(b_0) < C$ and $V^*(b_0) > V_{\rm opt}$}
	 	\State $\lambda_{\rm opt}=\lambda_0$
	 	\State $V_{\rm opt} = V^*(b_0)$
	\EndIf
\end{algorithmic}
\end{algorithm}
\begin{algorithm}[h]
\caption{Point-Based Value Iteration (PBVI) - ONLINE
\label{alg:PBVI_ONLINE}}
\begin{algorithmic}[1] 
\State $\tilde{\mathcal{B}}=\text{BELIEFS}$
\State Set $n=0$, $V_0=0$, $V_0^c=0$, $\lambda_0={\lambda}$, $\alpha_0={\alpha}$.
\State Define $\mathcal{L}(s,a)=r(s,a)-\lambda_0 c(s,a)$
	\For{$n=0,\dots$}	
		\State $V_{n+1} = \text{PERSEUS}(\tilde{\mathcal{B}},V_n)$
		\If{$|\sum_{b \in \tilde{\mathcal{B}}} V_{n+1}(b) / \sum_{b \in \tilde{\mathcal{B}}} V_n(b) -1|<\epsilon_V$}
			\If{$(V_{n+1}^c(b_0)-C)/C<\epsilon_c$}
				\State Break
			\EndIf
		\EndIf
		\State $\lambda_{n+1}=\max(0,\lambda_n+\alpha_n(V_{n+1}^c(b_0)-C))$
		\State Define $\mathcal{L}(s,a)=r(s,a)-\lambda_{n+1} c(s,a)$
	\EndFor
	\State $V^*(b_0)=V_{n+1}(b_0)$
	\State $V_c^*(b_0)=V_{n+1}^c(b_0)$
	\State $\lambda_{\rm opt}=\lambda_{n}$
	\State $V_{\rm opt} = V^*(b_0)$
\end{algorithmic}
\end{algorithm}

To find the optimal multiplier $\lambda_{\rm opt}$, we have to run the routine for different values of $\lambda$. ${\rm PERSEUS}$ performs a number of backup stages until some convergence criterion is met. At this point, we check if the constraint $V_c^*(b_0) < C$ is satisfied, update $\lambda_{\rm opt}$ if $V^*(b_0) > V_{\rm opt}$, and repeat the routine for different values of $\lambda$. These values of $\lambda$ can be sequentially selected from a sorted sequence or properly tuned at the end of the routine in a smarter fashion. However, in both cases we have to wait until convergence. To speed up the search for the optimal multiplier $\lambda_{\rm opt}$, we formulate an online version of Algorithm~\ref{alg:PBVI}, which is presented in Algorithm~\ref{alg:PBVI_ONLINE}. Here, $\lambda$ is properly tuned within the main loop of the routine according to a gradient descent technique~\cite{shalev2014understanding}
\footnote{Note that a gradient descent technique adjusts the parameter $\lambda$ after each iteration in the direction that would reduce the error on that iteration the most. The target here depends on the parameter $\lambda$, but if we ignore that dependence when we take the derivative, then what we get is a \mbox{semi-gradient} update~\cite{sutton1998reinforcement}.}: $\lambda_{n+1}=\max(0,\lambda_n+\alpha_n(V_{n+1}^c(b_0)-C))$, where the discount factor is $\alpha_n=\alpha_0/(n+1)$. Finally, given $\lambda_{n+1}$, we update the Lagrangian relaxation as $\mathcal{L}(s,a)=r(s,a)-\lambda_{n+1} c(s,a)$.  
In addition to the convergence criterion of the standard PBVI, we also consider the requirement $(V_{n+1}^c(b_0)-C)/C<\epsilon_c$.

\section{Numerical results}
\label{sec:Numerical_Results}

We set $\Delta_{\rm t}=10$~$\mu$s. We consider the following parameters: $\Theta=120^\circ$, $d_0=10$~m, $W_{\rm tot}=400$~MHz, $f_c=60$~GHz, $\xi=1$, $N_0=-174$~dBm, $S=10$, $\mathbb{E}[v]=20$~m/s, $P_{\text{DT}} \in \{10, 20, 30\}$~dBm, $T_{\text{DT}} \in \{1000, 2000, 3000\}$ (number of micro time-slots, i.e., $\{10, 20, 30\}$~ms), $T_{\text{BT}}=1$, $P_{\text{BT}}=P_{\rm min}$, where $P_{\rm min}$ follows from $\mathbb P_{\text{FA}}=\mathbb P_{\text{MD}}=\epsilon$ (specified below). Finally, we compare different sets $\tilde{\mathcal{B}}$. Let $D$ be the average duration of a transmission episode. The average rate (bit/s) and power (dBm) are
 computed as $V_{n+1}^r/D$ and $V_{n+1}^c/D$. 
 
 \begin{figure}[htb]
	\centering
	\includegraphics[width=.9\columnwidth]{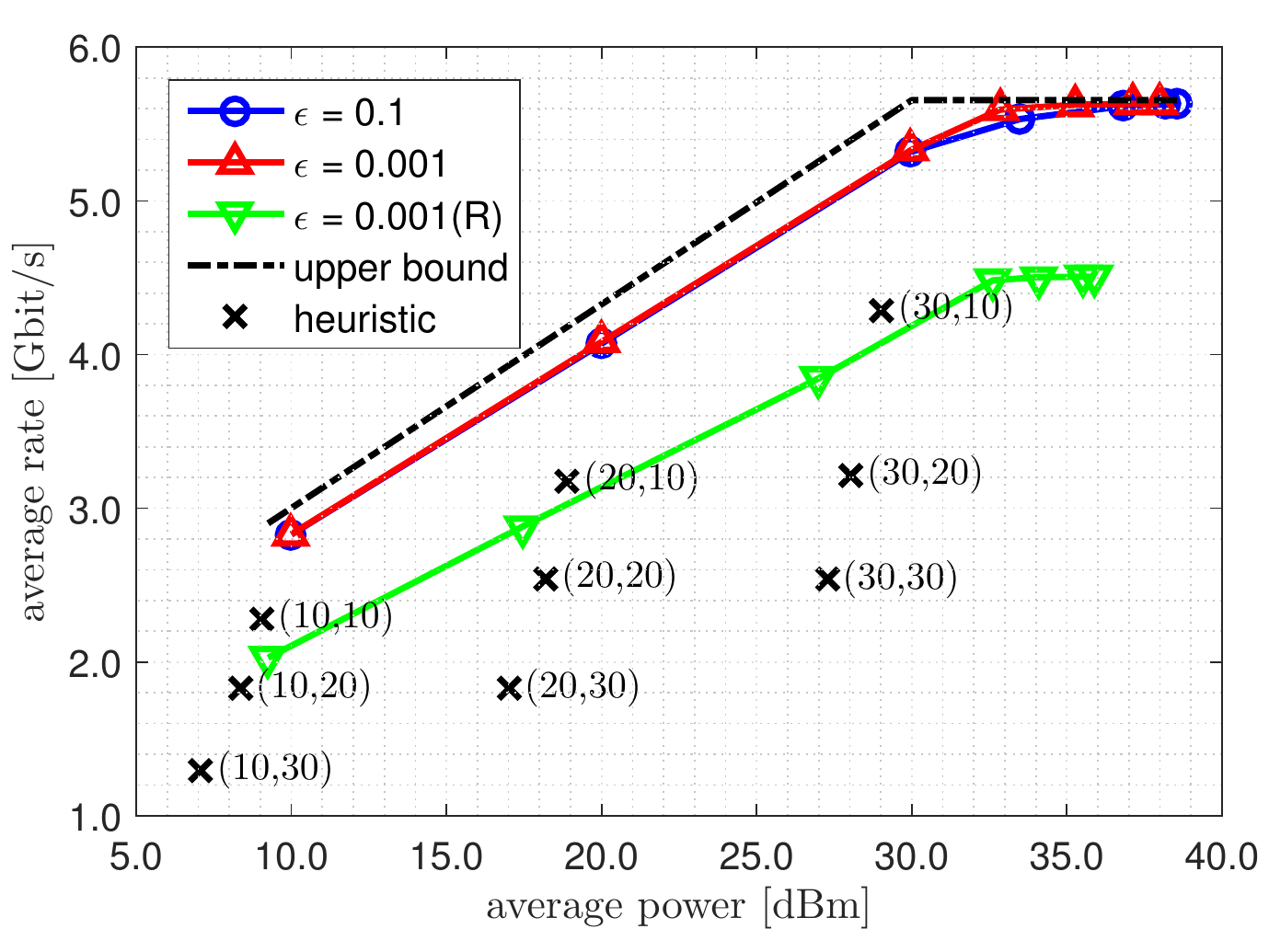}
	\caption{Rate and power as a function of $\lambda$ and comparison with the heuristic $\pi_{\rm H}$ for different pairs ($P_{\text{DT}}$, $T_{\text{DT}}$) (black crosses) and PBVI with points in $\tilde{\mathcal{B}}$ obtained by random sampling (R).}\label{figure:RATEvsPOWER_HEUR}
\end{figure}

The average rate and power as a function of $\lambda \in [10^5, 10^{11}]$ are plotted in \fig{figure:RATEvsPOWER_HEUR}, with $\epsilon=\{0.1,0.001 \}$. As $\lambda$ increases, both the average rate and power decrease, and the optimal policies are more conservative: given the current belief, the BS performs more directive DT over a smaller set $\mathcal{\hat S}$, using smaller values of $P_{\text{DT}}$ and $T_{\text{DT}}$. Conversely, as $\lambda$ decreases, both the average rate and power increse, and the optimal policies are more energy-demanding. Also, as $\lambda$ decreases, the impact of $\epsilon$ on the performance of the optimal policies is slightly more evident: given the average power, we can achieve a larger average rate with a smaller $\epsilon$, i.e., a larger $P_{\rm min}$, meaning that the optimal policies are more sensitive to the observation errors when performing BT than to the actual value of $P_{\rm min}$. Overall, the plot suggests that the performance of the optimal policies is quite robust to the observation errors when performing BT, whereas this is not true for any heuristic policy. The heuristic $\pi_{\rm H}$ works as follows: the BS performs BT with ($P_{\text{BT}}$, $T_{\text{BT}}$) over $\mathcal{\hat S}=\{ s \}$ (starting from $s=1$). If the MU replies with ACK, then the BS performs DT with ($P_{\text{DT}}$, $T_{\text{DT}}$) over $\mathcal{\hat S}$, until the first NACK. At this point, the BS scans the two states $s-1$ and $s+1$ in two different micro-slots. If the MU replies with ACK, then the BS repeats the scheme (starting from the new state).  
The heuristic $\pi_{\rm H}$ for different pairs ($P_{\text{DT}}$, $T_{\text{DT}}$) (black crosses) cannot achieve the performance of the optimal policies, which take advantage of the belief update mechanism and provide adaptive BT/DT procedures according to the current belief. The average rate and power for $\pi_{\rm H}$ are plotted in \fig{figure:RATEvsPOWER_HEUR} for $\epsilon=0.001$. The achievable rate drops significantly when considering $\epsilon=0.1$ (not shown in \fig{figure:RATEvsPOWER_HEUR}), since $\pi_{\rm H}$ does not provide any countermeasure to the observation errors when performing BT. An upper bound on the average rate is given when performing DT over the system state (i.e., assuming that the BS knows the position of the MU), thus achieving the maximum transmission rate without wasting power.
When the probability distributions collected in $\tilde{\mathcal{B}}$ are not very representative of the system dynamics history, then the performance of the optimal policies can greatly degrade, see \fig{figure:RATEvsPOWER_HEUR}, where the probability distributions are obtained by sampling random actions and observations, as suggested in~\cite{DBLP:journals/corr/abs-1109-2145}, and compared to the ones obtained by Algorithm~\ref{alg:fun_BELIEFS} with $W=S$. Here, the total number of beliefs collected in $\tilde{\mathcal{B}}$ is the same in the two cases. However, the performance of the optimal policies with beliefs obtained by Algorithm~\ref{alg:fun_BELIEFS} can provide an improvement in the achievable rate of $\sim 1$~Gbit/s ($1.15$~Gbit/s gap using an average power of $35$~dBm).

\begin{figure}[t]	
	\centering
	\includegraphics[width=.9\columnwidth]{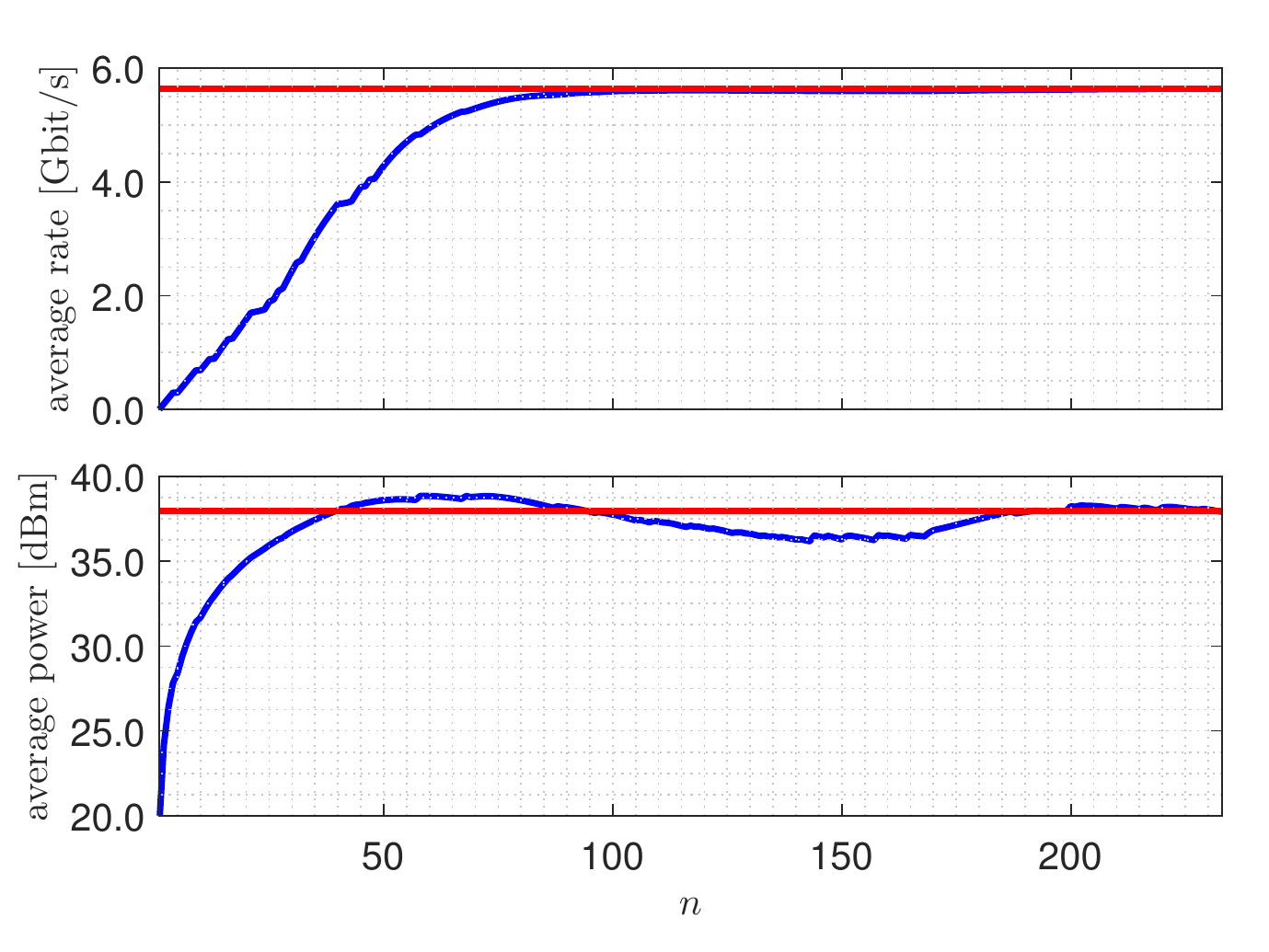}
	\caption{Rate and power as a function of $\lambda$, which is properly tuned according to $\lambda_{n+1}=\max(0,\lambda_n+\alpha_n(V_{n+1}^c(b_0)-C))$.}\label{figure:Fig_ADAPTIVE}
\end{figure}

An example of the outcome of Algorithm~\ref{alg:PBVI_ONLINE} is given in \fig{figure:Fig_ADAPTIVE}, starting from parameters $\lambda=0$ and $\alpha=100$. Here, we set the constraint $C$ 
(which corresponds to $V_c^*(b_0)$ for $\lambda=10^5$ in Algorithm~\ref{alg:PBVI}), and we achieve convergence of the rate to $\bar R$
, with $\lambda_{\rm opt}\simeq 10^5$. The average rate and power as a function of $n$ are plotted in \fig{figure:Fig_ADAPTIVE}, with $\bar R/D$ and $C/D$.


\section{Conclusions}
\label{sec:Conclusions}


In this paper, we have considered the transmission of data in a \mbox{mm-wave} vehicular network. There, base stations have to face the issue of beam training/realignment as the users move and, furthermore, have to concurrently decide the transmission power to use and how to balance transmission and beam training phases. This leads to a complex problem involving the concurrent estimation of user positions, to optimally decide upon: (i) alignment {\it vs} transmission activities, (ii) transmission power, number of beams and duration. Example numerical results reveal that optimal policies do provide a major advantage over heuristics, being able to maximize the transmission rate towards the mobile user, while showing robustness against beam-training errors, under a power constraint.    

\bibliography{bibliography}
\bibliographystyle{ieeetr}

\end{document}